\documentclass{article}

\usepackage[nonatbib]{neurips_data_2021}

\usepackage[utf8]{inputenc} 
\usepackage[T1]{fontenc}    
\usepackage{hyperref}       
\usepackage{url}            
\usepackage{booktabs}       
\usepackage{amsfonts}       
\usepackage{nicefrac}       
\usepackage{microtype}      
\usepackage{xcolor}         
\usepackage{graphicx}
\usepackage{epstopdf}
\usepackage{amsmath}
\usepackage{amssymb}
\usepackage{subcaption}
\usepackage{tabularx}
\usepackage[figuresright]{rotating}

\title{PSP: Million-level Protein Sequence Dataset for Protein Structure Prediction}

\author{%
  Sirui Liu,$^{1,}$\thanks{Equal contributions}\ $^{,}$\ \thanks{Corresponding Authors}\ \ \ Jun Zhang,$^{1,}$\footnotemark[1]\ \ Haotian Chu,$^{4,}$\footnotemark[1]\ \ Min Wang,$^{4}$\ \ Boxin Xue$^{1,2}$\\
  \textbf{Ningxi Ni,$^{4}$\ \ Jialiang Yu,$^{4}$\ \ Yuhao Xie,$^{1}$\ \ Zhenyu Chen,$^{2}$\ \ Mengyun Chen$^{4}$} \\ 
  \textbf{Yuan Liu,$^{2}$\ \ Piya Patra,$^{3}$\ \ Fan Xu,$^{6}$\ \ Jie Chen,$^{6}$\ \ Zidong Wang$^{4}$}\\
  \textbf{Lijiang Yang,$^{1,2}$\ \ Fan Yu,$^{4,}$\footnotemark[2]\ \ Lei Chen,$^{5,}$\footnotemark[2]\ \ Yi Qin Gao$^{1,2,3,}$\footnotemark[2]}\\
  \ $^{1}$Changping Laboratory\\
  \ $^{2}$Beijing National Laboratory for Molecular Sciences, \\College of Chemistry and Molecular Engineering, Peking University\\
  \ $^{3}$Institute of Systems and Physical Biology, Shenzhen Bay Laboratory\\
  \ $^{4}$Huawei Technologies Co., Ltd.\\
  \ $^{5}$The Hong Kong University of Science and Technology\\
  \ $^{6}$Peng Cheng Laboratory\
}

\begin{document}

\maketitle

\begin{abstract} \label{section:abstract}
  Proteins are essential component of human life  and their structures are important for function and mechanism analysis. Recent work has shown the potential of AI-driven methods for protein structure prediction. However, the development of new models is restricted by the lack of dataset and benchmark training procedure. To the best of our knowledge, the existing open source datasets are far less to satisfy the needs of modern  protein sequence-structure related research. To solve this problem, we present the first million-level protein structure prediction dataset with high coverage and diversity, named as PSP. This dataset consists of 570k true structure sequences (10TB) and 745k complementary distillation sequences (15TB). We provide in addition the benchmark training procedure for SOTA protein structure prediction model on this dataset. We validate the utility of this dataset for training by participating CAMEO contest in which our model won the first place. We hope our PSP dataset together with the training benchmark can enable a broader community of AI/biology researchers for AI-driven protein related research.

\end{abstract}

\section{Introduction}  \label{section:introduction}

Proteins are essential components of organisms and participate in most processes within cells. The 3D structure of the protein is arguably the most important property for understanding the function of the protein and how these functions are fulfilled. For a long time, the determination of protein’s tertiary structure given its sequence of amino acids in the polypeptide chain has been one of the popular topics for the scientific research community. Researchers deploy various experimental methods to determine protein's 3D structure, including X-ray crystallography\cite{ref1, ref2}, NMR (nuclear magnetic resonance)\cite{ref3, ref4} and cryo-EM\cite{ref5}. All these methods rely heavily on trial-and-error and expensive equipment. Computational tools for predicting structure and building 3D model directly from sequence are necessary, while traditional methods suffer from the problem of low accuracy \cite{ref6, ref7}. In recent years, the CASP (critical assessment of protein structure prediction) experiments\cite{ref8, ref9, ref10} have witnessed great improvement in  structure prediction quality by the application of powerful deep learning models \cite{trRosetta, ref12, ref13, ref14, ref15, ref16}. The most representative work is the AlphaFold2 \cite{alphafold} developed by DeepMind team which demonstrates prediction accuracy competitive with experiment in most cases. There are also other deep learning-based prediction methods, such as RaptorX \cite{ref12},  trRosetta \cite{trRosetta}, and RoseTTAFold \cite{ref18}.

Despite the remarkable advance in deep learning based protein structure prediction methods, details of these methods are still not clear for most researchers. Though some projects have open-sourced part of their work, the training or fine-tuning of these models are still privilege of a few teams for 2 reasons: 

1. There are not enough accessible training datasets in this area due to the high cost for their preparation. First, The labeling in this area depends on costly experiments. Only thousands of new proteins are resolved each year and the largest protein structure database, Protein Data Bank \cite{pdbdatabase}, now contains only 180k entries with tertiary structure. Besides labeling, most of structure prediction systems rely on traditional database searches to obtain the MSA (multiple sequence alignment) and template \cite{alphafold, ref18, ref19} which are also necessary for building training dataset for sequence-structure related research. Typically, generating a million-level protein dataset takes millions of CPU hour. Moreover, not all sequences are suitable for training and various defects require expertise knowledge to fix. The high cost results in the lack of open source dataset. To the best of our knowledge, so far, the most applicable training dataset accessible to the public is the trRosetta dataset released by Yang\&Baker Lab \cite{ref13}, which contains only 15k sequences and not enough for training.

2. The whole training  process is not clear. Most of the existing structure prediction systems are short of details about their training procedure. For example, DeepMind open sourced its inference code of AlphaFold2 but not the training code. Some description about the different stages of training is available but details about the accuracy in each stage of training is not clear\cite{alphafold}. 

To address these problems, we introduce in this work the first open source million-level protein structure prediction (PSP) dataset and the associated benchmark training procedure. Our main contributions are:

\begin{itemize}
\item The first million-level protein structure prediction database, which is the largest of its type with high coverage, diversity and quality.

\item The benchmark procedure for training SOTA protein prediction model, including the detailed hyper-parameters for training, training cost, and the expected accuracy after each stage of training, etc.

\item Improvements over the SOTA model, including new structure-related loss and new data sampling method for the initial training stage, as well as stabilized violation loss for the fine-tuning stage.
\end{itemize}

\section{Related work} \label{section:related_work}

In this section, we review the status of current training datasets and deep learning based techniques for predicting protein structure.

\paragraph{Training dataset} 
The trRosetta training dataset \cite{trRosetta} is a public dataset used to train neural networks for protein structure prediction which consists of 15051 protein sequences . Protein sequences with at least 40 residues and a resolution of less than 2.5\AA are selected. Each protein in the training set has at least 100 homologous sequences in MSA. The accuracy of the structural models trained on trRosetta appear to be close to AlphaFold2 at low computational cost while, due to the small number of entries, this dataset lacks diversity to train the AlphaFold2 at full scale. ProteinNet \cite{ProteinNet} is another standardized data set for deep learning of protein structure. It consists a series of datasets from V7 to V12 and cutoff dates for sequence and structure data from May, 1, 2006 to May, 1, 2016, which is too old as the number of entries with structure in PDB database increases more than 50\% in past 5 years.

\paragraph{Predicting protein structure using deep learning}

In recent years, deep learning techniques have greatly improved the accuracy of protein structure prediction. Some of the methods focus on prediction of 2D contacts/distances between residues. For example, Deep residual networks (RaptorX\cite{ref12}) and transformer-derived models(ESM-1b\cite{ref_esm1b}, TAPE\cite{ref_tape}, ProtTrans\cite{ref_prottrans}, etc) are designed to protein contact prediction. Distance prediction was also applied to replace binary contact prediction\cite{DistanceFolding}. The first Alphafold model\cite{ref16},  trRosetta \cite{trRosetta} and trRosettaX \cite{trRosettaX} are representatives models for this task. These methods typically use deep residual networks to predict protein structure related information, with relatively insufficient information on interactions between amino acid residues in long sequences.

In CASP14, AlphaFold2 \cite{alphafold} won the first place with obvious advantage. It directly predicts 3D atomic coordinates rather than 2D contacts or distances for a given protein using the multiple aligned sequences of primary amino acid sequences as input. It also replaced 2D convolution by an attention mechanism which can better represent long-range interactions between far residues of the sequences. It applies an end-to-end training approach that all parameters are optimized by back propagation of gradient from the generated 3D atomic coordinates through all network layers. Another end-to-end deep learning-based approach RoseTTAFold \cite{ref18} requires less memory for inference (8GB, less than 24GB of AlphaFold2 for proteins with 400 residues) and is close to AlphaFold2 in accuracy.

\section{Data Curation} \label{section:data_curation}

 In this section, we aim at building true and predicted structure datasets with additional curated information sufficient for protein sequence-structure prediction related tasks. According to rcsb PDB database statistics, by 2021 there are 52.9k clusters of 50\% similarity for experimentally solved structures, while the UniRef50 sequence database, which is clustered by the same similarity level, gives 48 million sequence clusters. The true structure database is relatively insufficient for large models, while the sequence data waits to be explored at multiple levels. Therefore to make better use of the structure database and the large sequence databases, we curated two main training datasets to obtain sufficient training data for protein structure prediction models. One is the true structure dataset curated from the experimentally resolved PDB structures, the other is the distillation dataset with sequences selected by maximizing diversity and quality and with structures predicted by AlphaFold. These two datasets together construct a million-level training set for PSP task and are used for benchmarking. To help evaluating model quality on the same basis, as part of the PSP dataset we provide a new validation set with no time overlap to the training set. Since the complete training set is relatively large for early model adjustment and hyperparameter tests, we additionally provide a PSP Lite dataset which is 1/10 the size of the full training set, but with higher MSA quality(see Appendix for details).

\subsection{True structure dataset}\label{section:pdb_data}

The Protein Data Bank (PDB) \cite{pdbdatabase} is the single global archive of experimentally 3D structure data of biological macro-molecules including sequence and 3D structures of proteins. However it lacks auxiliary information crucial for tasks like structure prediction,including similar solved protein structures(templates) and similar sequences aligned  to the query protein sequence (Multiple Sequence Alignments, MSAs). Due to complicated data contributors and conditions as well as its long time span, the PDB dataset contains samples with different naming and numbering strategies, modifications, and experimental quality, which calls for data cleaning. Therefore we make use of the PDB database by cleaning up the sequence-structure data and generating additional information of MSAs and templates.

We downloaded the structure data from PDB database with the version on October, 13, 2021, and obtained 170k multi-chain protein structure files. Since the inter-chain and intra-chain freedom is largely different, here we focus on providing dataset for single chain tasks, and separated the structure files into 600k single chains. The single chain structure data is then cleaned up so that alternative structures are removed and hetero atoms are processed (see Appendix for detailed description) to ensure one-vs-one sequence-structure correspondence and limit to 20 amino acid types. Since proteins are linear molecules with local secondary structures, sequential continuity is expected to help in local structure determination. To ensure structure continuity and residue correspondence, processed structure sequences are aligned to the original sequences with gaps filled in by UNK(signifying unknown residue). All sequence-structure pairs are filtered so that structures with high sequence identity or low resolution (>9 \AA) is excluded from the dataset as basic quality control. 

The aligned sequence is then used to generate MSA by searching against the UniClust30\cite{uniclust30} and Envdb database with MMseqs2\cite{mmseqs2} following the ColabFold protocol\cite{colabfold}. The  UniClust30 MSA profile is additionally applied to template searching against PDB70 database\cite{hhsuite3}. In this way, for each selected protein structure, we integrate a data bundle containing its sequence, aligned sequences (MSA), structure, and structural template information. Additionally, protein sequences are clustered by 40 percent mutual similarity, and the final total cluster number is 43.9k.

\begin{figure}
\centering
\begin{subfigure}{0.48\textwidth}
\flushleft
\includegraphics[width=6.2cm]{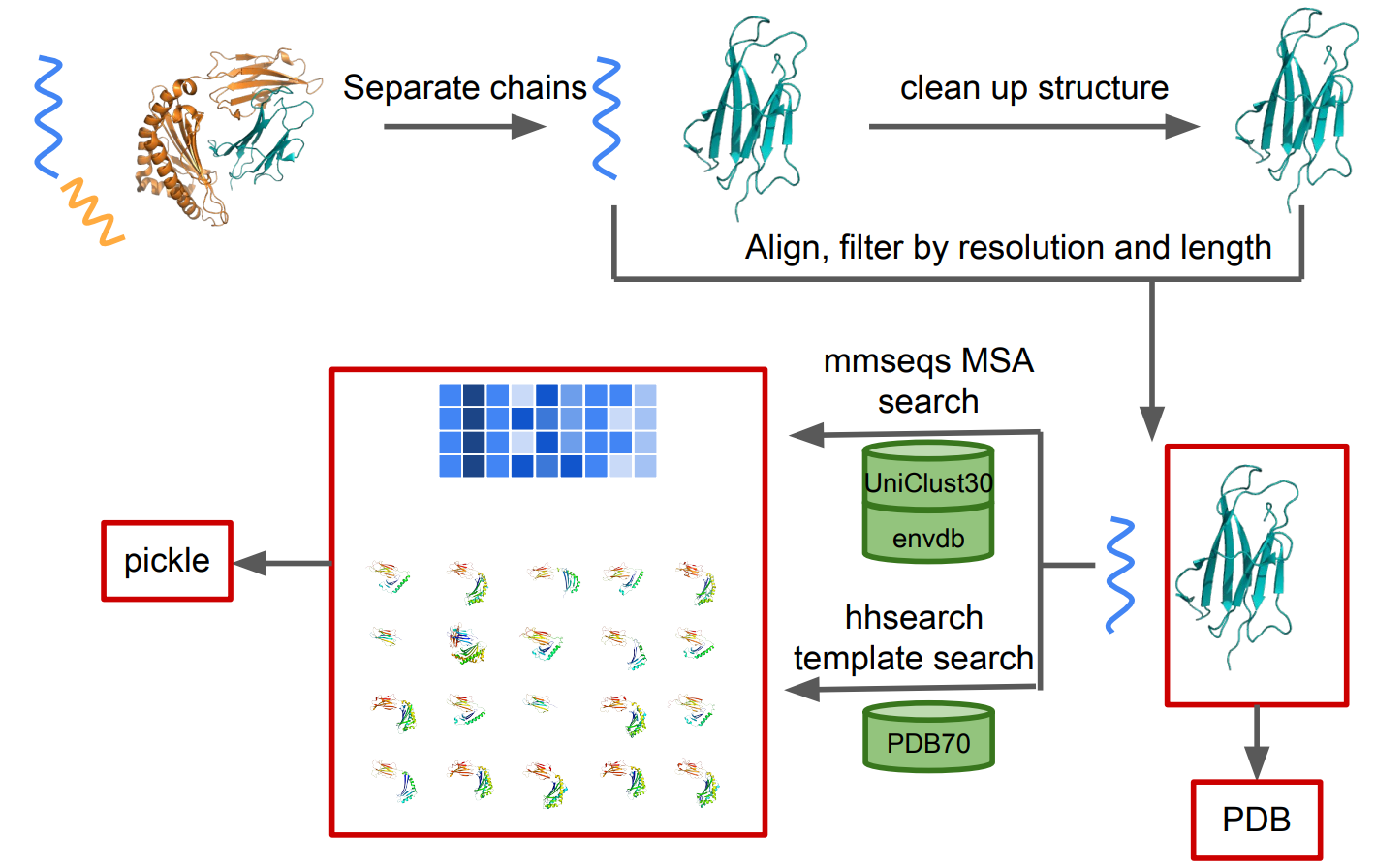}
\caption{}
\label{fig:true_scheme}
\end{subfigure}
\begin{subfigure}{0.48\textwidth}
\flushright
\includegraphics[width=6.8cm]{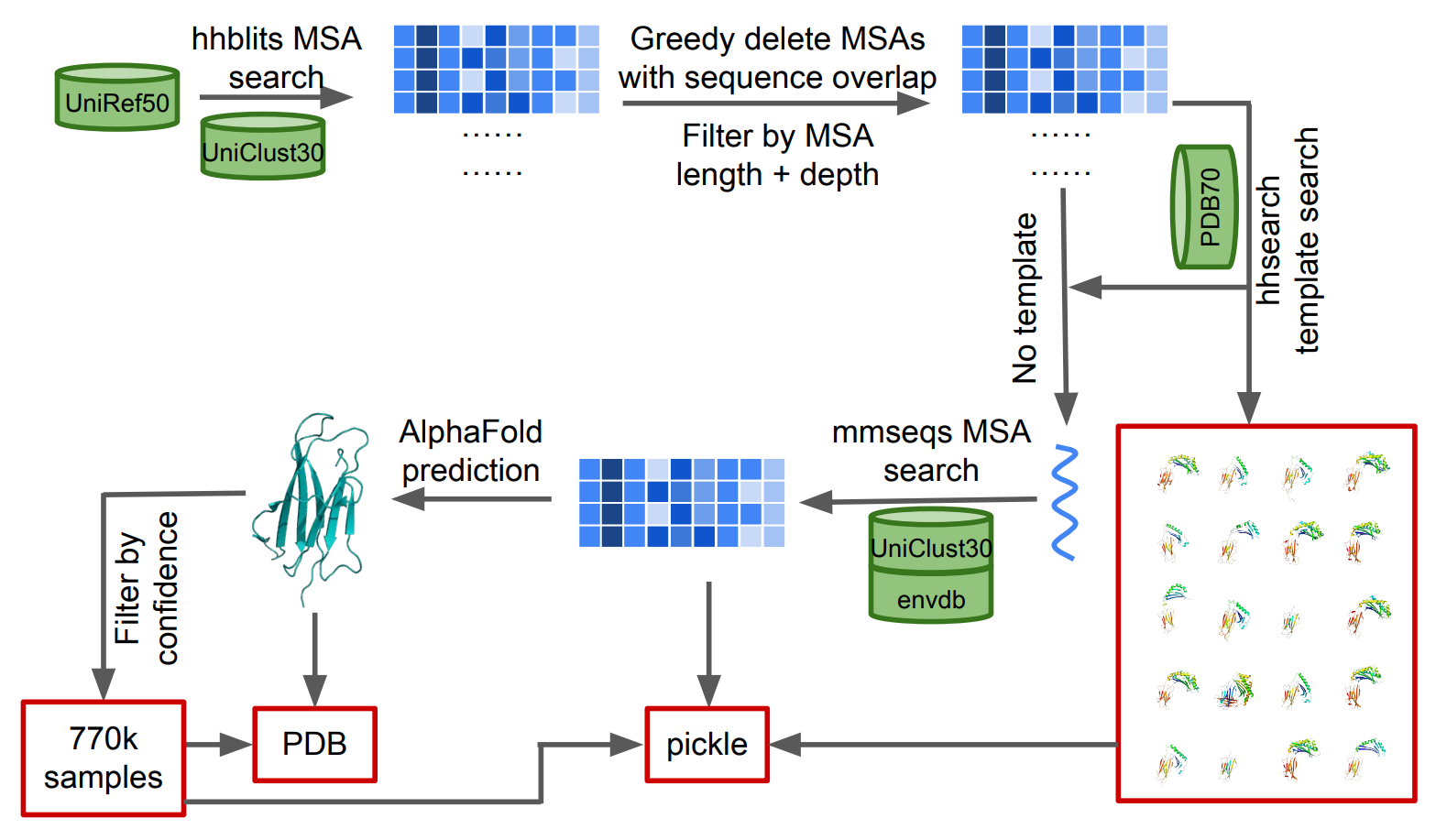}
\caption{}
\label{fig:distill_scheme}
\end{subfigure}
\caption{schematic figure for data pipeline of (a)true structure dataset and (b) distillation set.}
\end{figure}

\subsection{Distillation dataset} \label{section:distillation_dataset}

To make better use of the massive sequence data, we generated distillation dataset besides the true structure dataset. The samples in distillation set is aimed to be complementary to those with true structures, while balancing coverage and computational cost. The Uniref50\cite{uniref50} database contains in total 48 million sequences with mutual similarity lower than 50 percent. We previously have searched all Uniref50 sequences shorter than 1024 against the Uniclust30 cluster with HHblits\cite{hhsuite3}, and obtained a complete MSA set for the Uniref50 sequences.  We then filtered the sequences by their similarity, length, and MSA quality with a protocol similar to AlphFold2. Based on the UniRef50 MSAs we greedily removed all sequences that can be found in other MSAs to make the remaining sequences as diversed as possible. Since structure prediction of short sequences is a relatively simple task, sequences with lengths shorter than 200 are then filtered out. AlphaFold2 has shown that its structure prediction accuracy is related to MSA depth(number of sequences in a MSA), we therefore selected sequences with MSA depth above 200 after identity and coverage filter(performed using HHfilter\cite{hhsuite3}) to ensure structure prediction quality. We acquired 4.1M sequences after this procedure. 

To make the distillation dataset complementary to the true structure dataset, we selected sequence by their similarity to the PDB70 dataset. Only sequences with highest sum of probability(sum\_probs) over length in HHsearch\cite{hhsuite3} lower than 0.4 (see Appendix for detailed illustration) are selected to ensure that sequences in distillation set are dissimilar to those with solved structures. Such a procedure yields overall 856k distilled protein sequences. In this way we intentionally build the distillation set with diversity both within the dataset and in comparison to the true structure dataset.

To generate MSA and predicted structures, all the sequences ready for distillation are searched against Uniref30 and Envdb dataset with MMseqs2 following the same protocol described in \ref{section:pdb_data}. Since the selected sequences are distinguished from the sequences in true structure dataset, templates are not expected to play important roles in structure prediction. The MSAs are fed into AlphaFold2 model\_1 without template and predicted structures are generated in batch. All predicted structures are then filtered with plDDT score, and those with AlphaFold2 violation loss higher than zero are further relaxed by openMM\cite{openmm} following the standard AlphaFold2 protocol to improve data sanity. For distillation set, we obtain 760k data bundles with protein sequence, MSA, template, and predicted structure.

\subsection{Validation set}\label{section:validation_dataset}

In training and evaluation, we used two validation sets. One is the widely recognized CASP14 dataset for which we generated MSAs and templates following the same protocol as \ref{section:pdb_data}, and the other is a new validation dataset generated from publicly available data and is a combination of two datasets: The first is the CAMEO targets from Oct 16, 2021 to February 12, 2022, the second one is composed of new single clusters from the official PDB cluster by 40 percent identity between Oct 13, 2021 and March 15, 2022. The second part is selected so that all the targets are unique and are not similar to the training dataset. After filtering samples with lengths shorter than 1536 for convenience of prediction, the combination of the two datasets gives 513 non-overlapping samples in total. All sequences in new validation set are generated with ground truth labels after Oct 13, 2021, strictly after the date of all training samples, template and sequence searching databases, therefore could strictly prevent data leak in validation/testing.


\subsection{Dataset summary}\label{section:dataset_summmary}

To summarize, we composed a true structure dataset with 570k protein sequences and corresponding experimental structures, MSAs, and structural templates, and a complementary distillation set with 760k protein sequences and corresponding predicted structures, MSAs, and templates. The two dataset are organized in the same way and can be easily integrated, making a total set of over a million protein samples for training of large protein models. Additionally, we organized a new validation set of 513 samples free of data leaking.

\section{Benchmark} \label{section:benchmark}

To assess the utility of our dataset, we compared the training of SOTA model on our PSP dataset and on trRosseta dataset. We also adapted the original model to  train our dataset. The model is used to participate in the CAMEO contest and won the first place, which validates the training effect on our dataset.

\subsection{Experiment setup} \label{section:experiment_setup}

We trained the AlphaFold2 model \cite{alphafold} in all experiments as it is the current SOTA in this field. We took the TM-score \cite{tmscore} on 87 CASP14 sequences\cite{ref10} as the monitor of accuracy. The TM-score (template modeling score) is used to assess the topological similarity between protein target structure and predicted structure. A value 1 means exactly the same, and <0.17 means totally irrelevant \cite{tmscore2}. The original model yields a 0.86 average TM-score. To make use of CASP14 test set, we excluded all the sequences with structure release date after May 13, 2020, as mentioned in \cite{alphafold}, to prevent data leakage since the CASP14 started after this date. We also tested the accuracy on our validation set\ref{section:validation_dataset} on which original model obtains a TM-score of 0.82. 

We trained the model with MindSpore AI framework\cite{mindspore}. The MindSpore framework supports complex control flow syntax, highly efficient memory management and auto mixed precision computing which are necessary for training this model. We trained the model with a batch size 1 on 128*Ascend 910. With this setting, 75k steps of training are equivalent to training with 9.6M samples, which conforms original protocol. Due to the difference in hardware, software and training dataset, we adopted different hyper-parameters for training while introducing several adaptations to the original training protocol. 

\subsection{Datasets comparison} \label{section:datasets_comparison}

We first compared the training accuracy using our PSP dataset and that using trRosetta dataset. trRosetta dataset \cite{ref13} is an applicable and common training dataset for protein structure prediction task open to public. It contains 15k sequences in total and is also equipped with MSA and template information. For this comparison, we adopted a cosine decay (from 1e-3 to 5e-4) learning rate with 1k steps of warm-up and gradient clipping with a clipping value 1.0 by the global norm after the gradient aggregation. The data samples used for training is 1.5 million for both, resulted in 100 epochs of trRosetta and only 1.5 epochs for our dataset. 

\begin{figure}
\centering
\begin{subfigure}{0.48\textwidth}
\flushleft
\includegraphics[width=7cm]{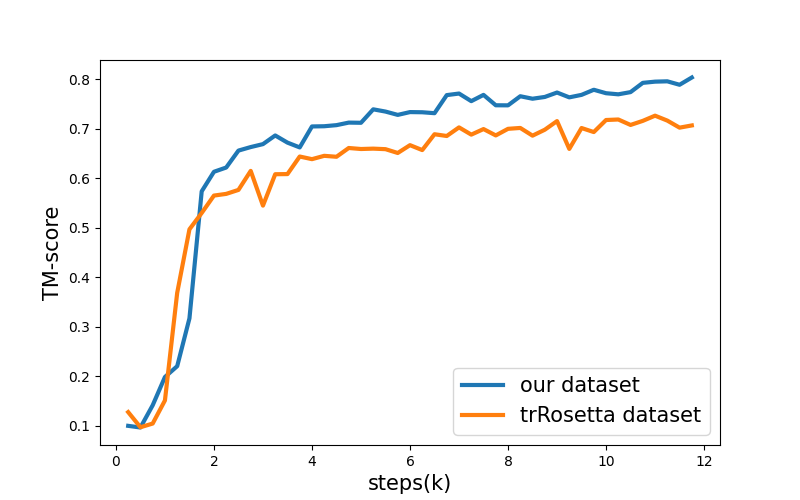}
\caption{}
\label{fig:samples}
\end{subfigure}
\begin{subfigure}{0.48\textwidth}
\flushright
\includegraphics[width=7cm]{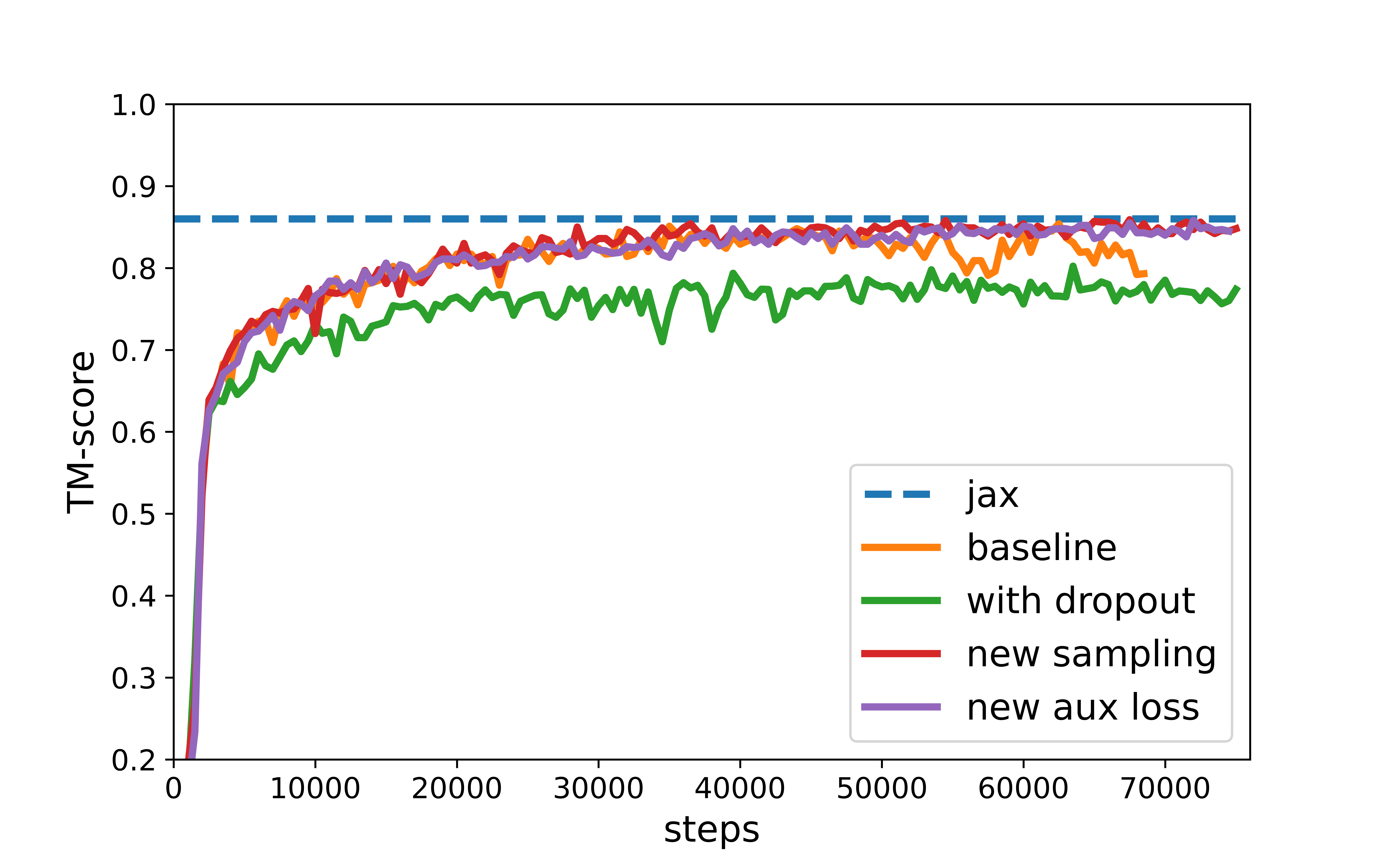}
\caption{}
\label{fig:init_training_accuracy}
\end{subfigure}
\caption{TM-score on CASP14 test set for training with different datasets(a) and  for different setup of training(b).}
\end{figure}

 As presented in Figure \ref{fig:samples}, the model takes less time to converge on the trRosetta dataset, the TM-score in CASP14 when training with trRosetta set increases more quickly in the early stage of training, but converges to 0.7, significantly lower than the training with our PSP dataset, which converges to 0.8.

\subsection{Training Performance} \label{section:training_performance}

We adopt the 2-stage training process (initial training + fine-tuning) and several modifications has been made to each stage. 

\subsubsection{Initial training} \label{section:init_training}

\paragraph{Baseline}

We take the initial training of the original model as the baseline for this section. As hyper-parameters, we adopted a cosine decay (from 1e-3 to 5e-4) learning rate and applied the gradient clip by global norm with clipping value 1.0. We discarded all dropout in this model since our dataset is more diversified. The dropout is a normalization technique which prevents over-fitting and is not unnecessary for training on our dataset. In experiment, the model converges more efficiently to higher accuracy without dropout \ref{fig:init_training_accuracy}: training with dropout takes 20k steps to attain 0.75 and converges to 0.79, while without dropout takes 10k steps to attain 0.75 and converges to 0.84 at 45k steps. However, the training is unstable and the accuracy drops to under 0.8 after 50k steps. We therefore introduced other techniques to improve the training on our dataset.

\paragraph{Data sampling method} 

\begin{figure}
\centering
\begin{subfigure}{0.58\textwidth}
\flushleft
\includegraphics[width=8cm]{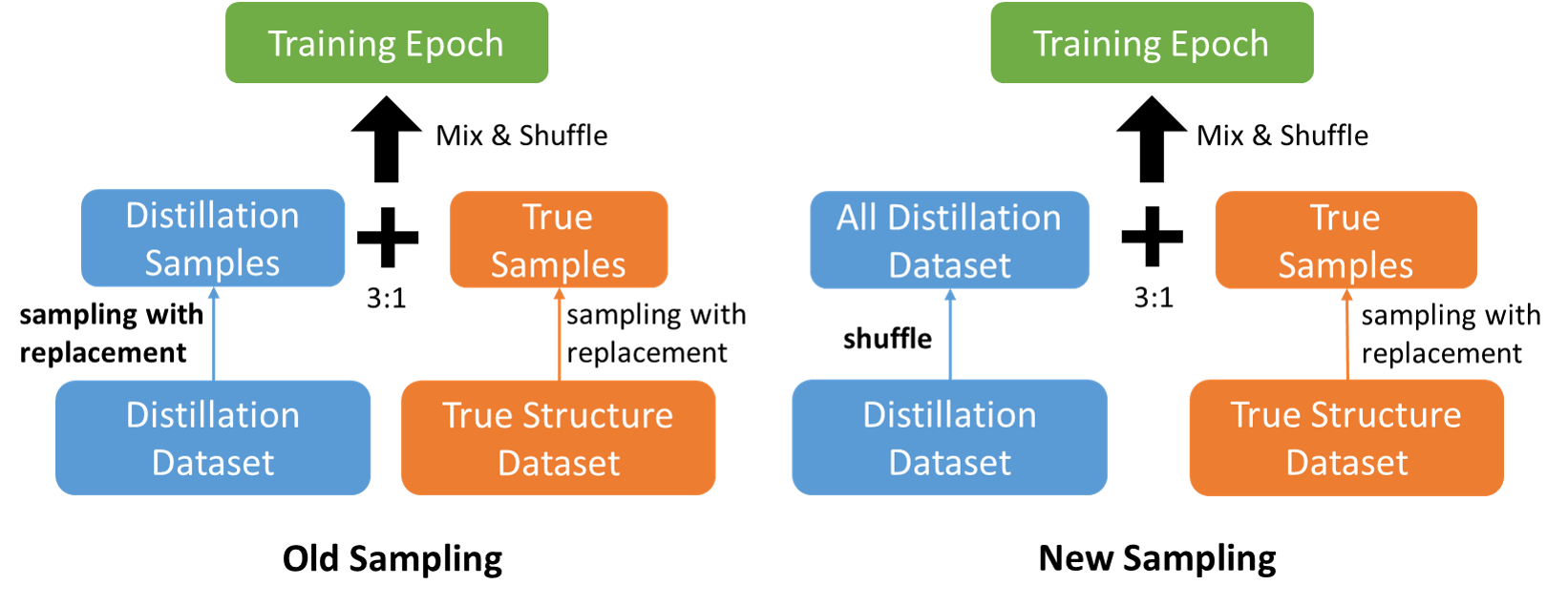}
\caption{}
\label{fig:sampling_method}
\end{subfigure}
\begin{subfigure}{0.38\textwidth}
\flushright
\includegraphics[width=5.5cm]{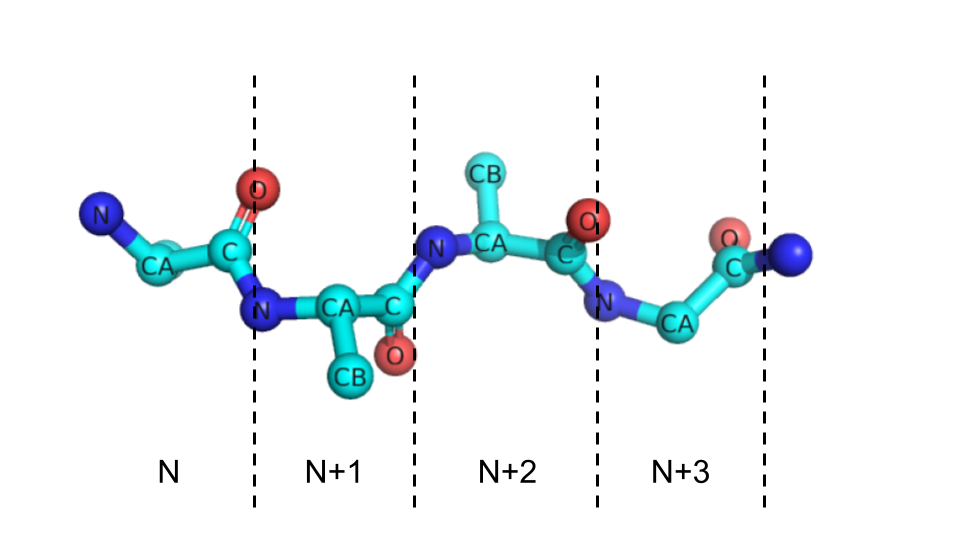}
\caption{}
\label{fig:proteins}
\end{subfigure}
\caption{Comparison between different sampling method(a): baseline sample method(left) and our new sampling method(right); Demo of protein chain of 4 amino acid residues(b)}
\end{figure}

As mentioned in \ref{section:data_curation}, we’ve curated a diversified subset of UniRef50 sequences for the 760k distillation dataset, while the true structure set contains effectively 43k clusters of protein sequences, which is far less diversed than the former. For the baseline, each epoch samples with replacement from the whole distillation dataset. This sampling method ensures that the data is different in each epoch, but does not guarantee that all data pass the training in early stages. As a result, there are still new samples unseen by the model at the last 15k steps, which may destabilize the training. Based on this analysis, we proposed a new sampling method, as shown in figure \ref{fig:sampling_method}. This new method takes the traditional notion of epoch for the distillation dataset, and each epoch of training data  consists of all distillation data and true structure data sampled with replacement from whole true structure dataset. This sampling method ensures all distillation data pass the training for about every 10k steps to make full use of this dataset. As shown in figure \ref{fig:init_training_accuracy} (new sampling), the model converges faster to a final accuracy 0.86, equivalent to the released checkpoint of DeepMind which is trained with 2 extra stages of fine-tuning.

\paragraph{Loss terms} 

We also introduced more structure-related constraints to improve the training on our dataset. In baseline, the structure-related constraints consist of the final atom position loss and the intermediate output position loss $\mathcal{L}_{aux}$ \cite{alphafold}. The aux loss is composed of the Frame Align Point Error loss and the angle norm loss. The FAPE loss calculate the atom position error in each local frame of residue. Because of the unstable gradient in recycling of structure module, the FAPE loss term $\mathcal{L}_{FAPE\_C_{\alpha}}$  of $\mathcal{L}_{aux}$ considers only the position of $C_\alpha$ atom (CA in figure \ref{fig:proteins}) but ignores all other atoms in the residue(ref eq \ref{eq:fape_ca}). This simplification $\mathcal{L}_{FAPE\_C_{\alpha}}$ does not have regularization effect on the orientation of the residues.

\begin{small}
\begin{align}
    \mathcal{L}_{FAPE\_C_{\alpha}} & =  {1 \over {N_{frames} * N_{C_{\alpha}}}} \sum\limits_{j=1}^{N_{frames}} \sum\limits_{i =1} ^{ N_{C_{\alpha}}} (\sqrt{\|\overrightarrow{x}_{ij} - \overrightarrow{x}_{ij}^{True}\|^{2} + \sigma})  \label{eq:fape_ca}
\end{align}
\end{small}

As three points can fix a rigid body in 3D space, given the hypothesis of baseline that the backbone of each residue is considered as a rigid body, we propose to add the point error of the C and N atom, atoms connected directly to $C_\alpha$ in all residues (see figure \ref{fig:proteins}), $\mathcal{L}_{FAPE\_C_{\alpha}CN}$ (eq \ref{eq:fape_cacn}), to the auxiliary loss to indirectly introduce the orientation information to this loss. We expect that, with more structure-related information, the training would be more stable. As shown in figure \ref{fig:init_training_accuracy} (new aux loss), with 
the added constraints, the training converges to 0.86 without decrease of accuracy at the later phase of training, which confirms the validity of this improvement.

\begin{small}
\begin{align}
    \mathcal{L}_{FAPE\_C_{\alpha}CN} & =  {1 \over {N_{frames} * N_{C_{\alpha} + C + N}}} \sum\limits_{j=1}^{N_{frames}} \sum\limits_{i =1} ^{ N_{C_{\alpha}  + C + N}} (\sqrt{\|\overrightarrow{x}_{ij} - \overrightarrow{x}_{ij}^{True}\|^{2} + \sigma})  \label{eq:fape_cacn}
\end{align}
\end{small}

\subsubsection{Fine-tuning} \label{section:finetuning}

The TM-score measures the accuracy of backbone atoms in protein. Prediction with model trained from initial training can have problems associated with the side-chain atoms, for example, the clash of atoms and wrong peptide bond lengths and angles (too large/small). These defects will affect downstream tasks of structure prediction. The original model used the violation loss $\mathcal{L}_{viol}$ \ref{eq:violloss} to regularize these defects. This loss is composed of 3 terms, with the first 2 terms $\mathcal{L}_{bondlength}$ and $\mathcal{L}_{bondangle}$ regularizing peptide bond length \& angle. The last term $\mathcal{L}_{clash}$ regularizes the clash between atoms within residues and clashes between residues\cite{ref18}.

\begin{small}
\begin{align}
    \mathcal{L}_{viol} & = \mathcal{L}_{bondlength} + \mathcal{L}_{bondangle} + \mathcal{L}_{clash}  \label{eq:violloss}
\end{align}
\end{small}

Since our initial training has reached the same TM-score level as DeepMind checkpoint, during fine-tuning, we focused on minimizing structure violation while keeping the other losses stable. Therefore, we used a smaller learning rate(1e-4) and extended the sequence crop length from 256 to 384 to render the model the capability of structure prediction on longer sequences. We kept $max\_msa\_clusters$ to be 128 which accelerates $20\%$ the training without defecting the model’s accuracy according to our experiments.

\begin{figure}
\centering
\begin{subfigure}{0.48\textwidth}
\centering
\includegraphics[width=7cm]{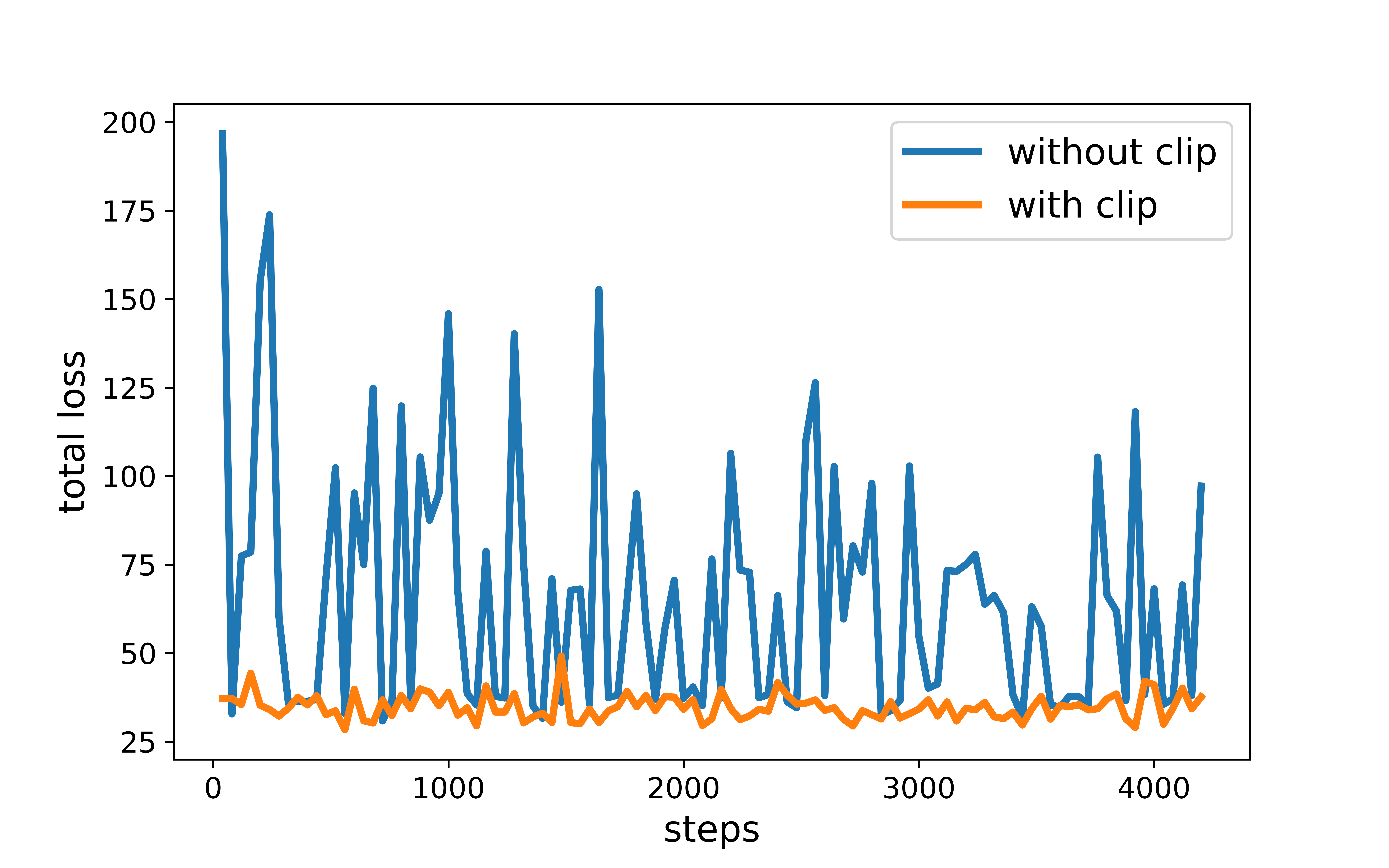}
\caption{}
\label{fig:finetune_loss}
\end{subfigure}
\begin{subfigure}{0.48\textwidth}
\centering
\includegraphics[width=7cm]{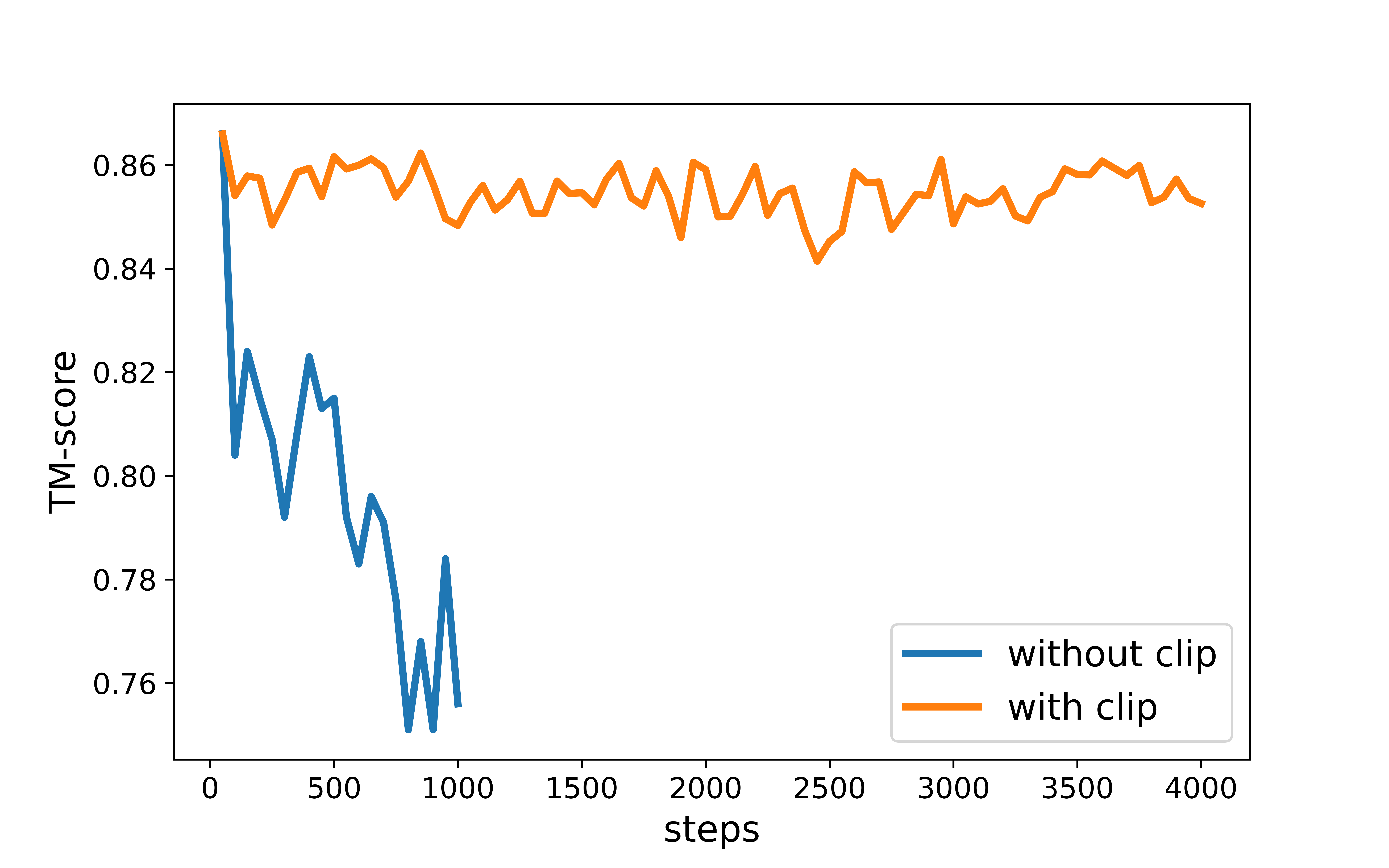}
\caption{}
\label{fig:finetune_tmscore}
\end{subfigure}
\caption{Total loss curves (averaged over 40 steps) (a) and TM-score on CASP14 test set (b) for fine-tuning with or without clip for violation loss term.}
\label{fig:finetune}
\end{figure}

\begin{figure}
\centering
\begin{subfigure}{0.48\textwidth}
\centering
\includegraphics[width=6cm]{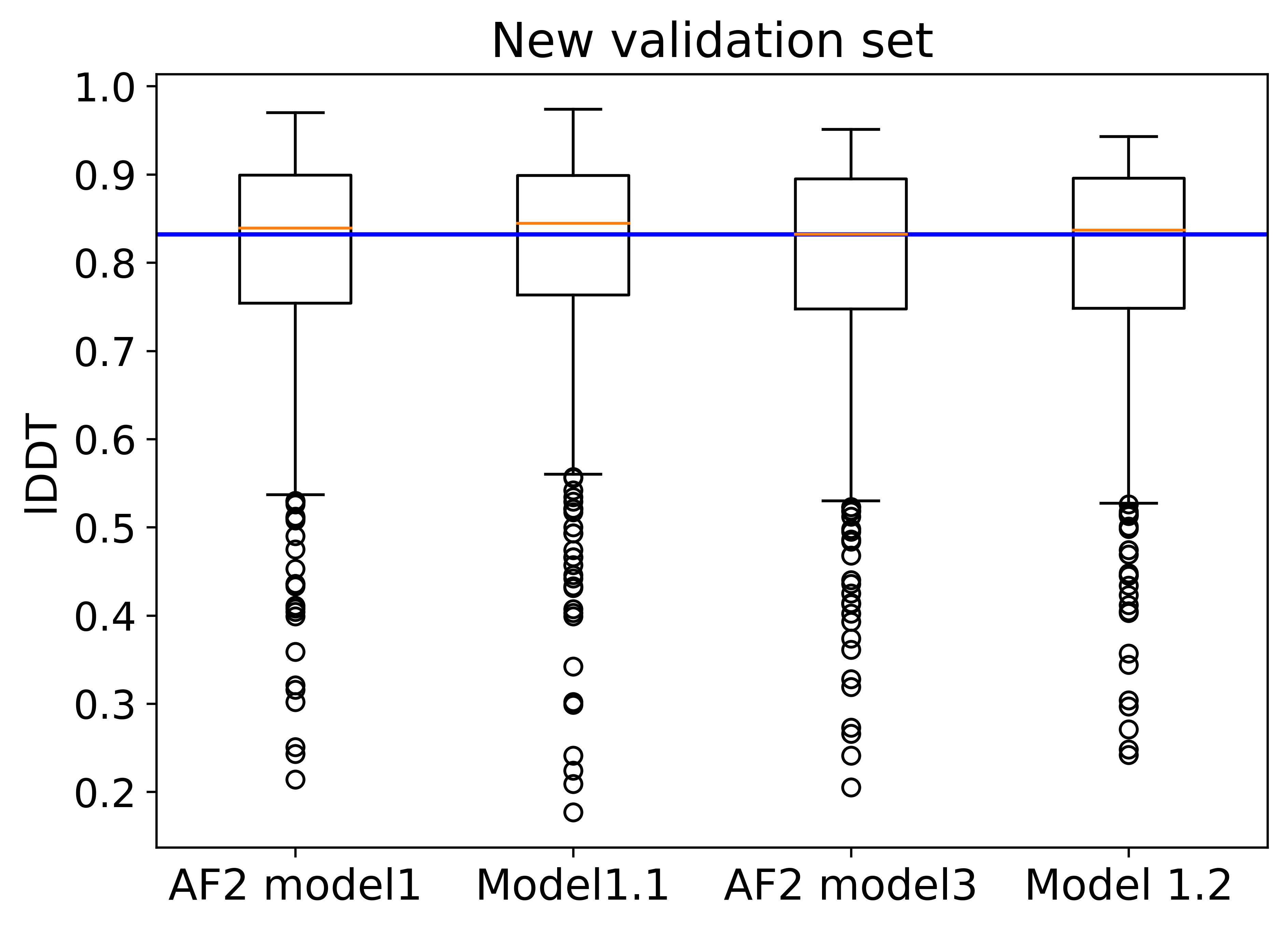}
\caption{}
\end{subfigure}
\begin{subfigure}{0.48\textwidth}
\centering
\includegraphics[width=6cm]{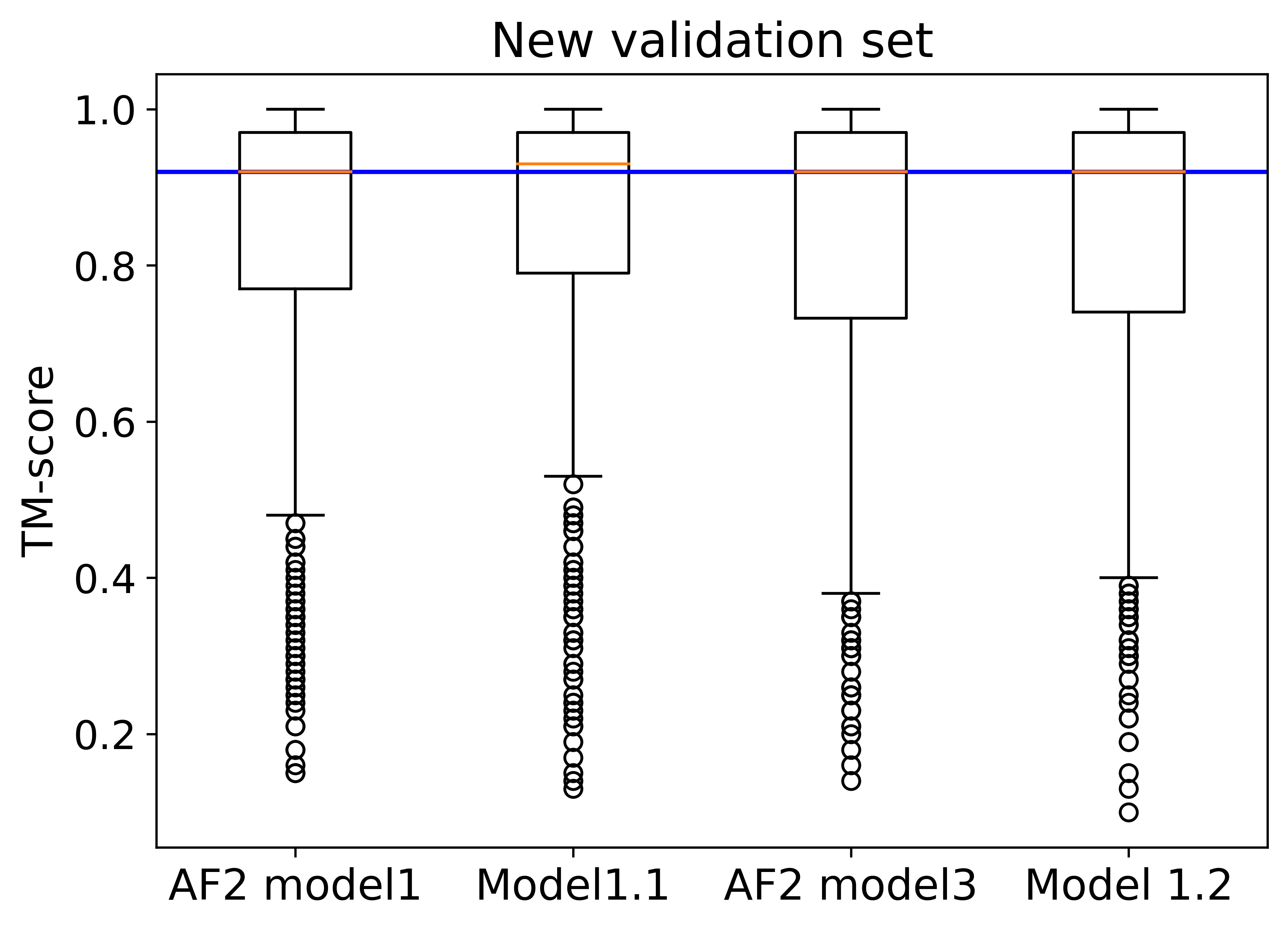}
\caption{}
\end{subfigure}
\caption{ Comparison of accuracy of our models(model 1.1 for model with template and 1.2 for without template) and AF2 model1(with template) and model3(without template), measured by lDDT(a) and TM-score(b) on new validation set.}
\label{fig:new_val_score}
\end{figure}

We found that with dropout, the TM-score on CASP14 test set dropped rapidly during fine-tuning, independent of the choice of training hyper-parameters. Moreover, in this setting the violation loss cannot over-fit on a single data even after thousands of iterations. Based on these experiments, we also turned off dropout in the fine-tuning stage.

Further experiments showed that, with dropout off, violation loss for part of the training data is two orders of magnitude larger than the others. The diversity of distillation dataset is useful to improve the model's generality, but the model did not learn enough violation information from these samples during the initial training. These abnormal loss crashed the training and resulted in a drop of TM-score. The main source of these abnormal loss was found to be due to the violation loss term $\mathcal{L} _{clash}$. To resolve the crash, we next cliped this term to 5.0 as in equation \ref{eq:new_violloss}.

\begin{small}
\begin{align}
    \mathcal{L}_{viol} & = \mathcal{L}_{bondlength} + \mathcal{L}_{bondangle} + clip(\mathcal{L}_{clash}, 5.0) \label{eq:new_violloss}
\end{align}
\end{small}

As shown in figure \ref{fig:finetune}, with this clip, the training loss is stabilized \ref{fig:finetune_loss} and the TM-score on CASP14 test set maintains to be 0.86 \ref{fig:finetune_tmscore}. Further analysis shows that the violations drops to nearly 0 after thousands of steps. To evaluate the accuracy of side-chain atom positions, we compared the lDDT (local Distance Difference Test, 0 as totally different and 1 as exactly the same) of predicted structured in which all atom are taken into account. On our validation set, our model yields a score of 0.80, the same as DeepMind checkpoint. More detailed comparison of scores on our new validation set can be found in figure \ref{fig:new_val_score}, with the same setting(with or without template), our models are all at least equivalent to the original model.

\subsection{Cameo Assessment} \label{section:cameo_ssessment}

The Continuous Automated Model Evaluation (CAMEO) \cite{cameo1, cameo2, cameo3, cameo4} complements the biennial CASP experiments with a fully automated blind assessment of the 3D protein prediction server based on weekly pre-releases of the sequences with structures, which is subsequently published in PDB release. We trained the original model with 2 different settings and the main difference between them is whether template are adopted (model 1.1 with template and 1.2 without template). We also tried the scoring method of Sergey et al. \cite{sergey_score} (quality scoring). During the 1-month evaluation cycle of CAMEO from March, 25, 2022 to April, 16, 2022, our models ranked first (model 1.1), second (scoring model) and fourth (model 1.2) respectively, surpassing all other existing models. More details about our models for the contest and the results can be found in  the Appendix.

\section{Limitations and Future Works } \label{section:future_work}

As a million-level dataset, our PSP dataset is prioritized on sequence diversity and aims at expanding beyond the PDB data. In this case, the size of distillation set could be expanded with loosening of diversity and natural increase in sequence database. In MSA/template searching, we noticed that most of the meta-genomics databases against which we searched are constructed before 2020, and the collection and construction of new sequence databases would be necessary to increase MSA diversity and add more evolutionary signals to the dataset. The effectiveness of out PSP dataset is demonstrated by training the SOTA protein structure prediction model from scratch. However, it should be noted that the application of the our datasets could be expanded to tasks beyond protein structure prediction, such as  protein sequence/MSA pre-training models, sequence and structure based protein design, sequence and structure generation, etc. We are looking forward to further explorations in these area.

\section{Conclusion} \label{section:conclusion}

Deep learning based protein structure prediction models have growing impact in protein structure and sequence similarity related research, while the training of such models are still restricted by the lack of training dataset and benchmark. We address this problem by presenting the first million-sample protein structure prediction dataset. With 1.3M sequences with true or distilled structure, this dataset has much higher coverage, diversity and depth compared to the existing open dataset. We provide in addition the benchmark of training SOTA protein structure prediction model on this dataset. Several improvements have been made to the original model to make better use of our dataset. We validated the utility of our dataset by participating CAMEO contest in which our model won the first place. We give details of the training procedure to researchers as a clear benchmark about training protein structure prediction models. All our work including the dataset is open to public, we hope that our work can enable a broader community of AI/biology researchers for AI driven protein related research.

\section*{Acknowledgments and Disclosure of Funding}\nonumber

This work was supported by National Natural Science Foundation of China (22050003, 92053202, 21821004, 21873007) and National Key R\&D Program of China (Grant Number 2021ZD0110400). We greatly acknowledge the support of MindSpore.

\section*{Data and Code Availability}\nonumber

The training code can be found at the repository of \href{https://gitee.com/mindspore/mindscience/tree/master/MindSPONGE}{MindSPONGE}. The full dataset will be hosted on a \href{http://ftp.cbi.pku.edu.cn/pub/psp/}{public server} in Peking University/Changping Laboratory. The PSP dataset will be publicly released very soon and will be fully accessible. The dataset is under \href{https://creativecommons.org/licenses/by/4.0/}{Creative Commons 4.0 license} and code is under \href{https://www.apache.org/licenses/LICENSE-2.0.html}{Apache 2.0 license}.

\newpage
\appendix

\section{Appendix}

\subsection{Details for Data Curation}

\subsubsection{Database in use}\label{section:sup_database}

Details of database used in data curation can be found in table\ref{table:database}, all of them are  available under Creative Commons 4.0 license. Our dataset is under the same license.

\subsubsection{True structure dataset cleaning}\label{true_cleanup}

We downloaded the raw protein structure files as pdb format according to the single chain structure list officially provided by rcsb PDB website\cite{pdbdatabase}. The list is the newest version on Oct 13, 2021. We used pymol to fetch and separate single chains. The multi-chain fasta files are downloaded from rcsb PDB database directly according to the same list, and are then divided into single chain fasta files. Note that the chain names in pdb files might be different from that in cif format which can be checked in the downloaded fasta files, and we adopted the [Auth] chain names. 

The pdb files from PDB database are diversed in naming, alternative conformations, modifications, etc, therefore needs further cleaning. The following cleaning protocol is carried out over all pdb files. Only the lines with ' '(space) or 'A' in column 17 and ' ' in column 27 are preserved to avoid alternative conformations. We also substituted phosphorylated residues to their natural amino acid correspondence: For lines of 'PYR' and 'Y1P' residues, lines with atom 'P' ,'O1P', 'O2P' and 'O3P' are deleted, 'HETATM' are substituted as 'ATOM' in the beginning of the line, and the residue name are changed to 'TYR'. The same line editions are performed for 'SEP' and 'S1P' to 'SER' and with 'TPO' and 'T1P' to 'THR'. The residue name 'MSE' are modified as 'MET' and atom name 'SE' are modified as 'SD'. For other lines which begins with 'HETATM', only lines with atom names of 'CA', 'C', 'N', 'O' are preserved. For residue 'ARG', we check the lines with atom name 'NH1' and 'NH2' to assure the distance of NH1 and CD are smaller than the distance of NH2 and CD. 

To set up structure related labels, residue-level correspondence is needed for the fasta file and pdb file. The cleaned pdb files are transformed to sequences (new fasta files) using pdb2fasta\cite{pdb2fasta} package. New fasta files are then aligned to the raw fasta files (fasta files downloaded directly from PDB database) using Biopython\cite{biopython}. This step aims to make sure that the cleaned protein structures have correct correspondence to the original sequences. The original sequence is further trimmed according to the alignment and the final sequences are saved as the processed fasta files. 

The index of residues in cleaned protein structure files are renumbered starting from 1 according to the processed fasta files. We then carried out a double-check to make sure that the index of last residue is the same as the sequence length. The current version of pdb files allow gap, while fasta file is continuous, this is because we hope the model to learn the importance of sequence continuity and the meaning of neighboring residues in structure prediction. To make structure loss calculations easier, we filled gaps with UNK so that both pdb and fasta are continuous in residues and the missing ground truth in structure can still be easily masked. In this way the sequence and structure files are cleaned up, and fasta-pdb residue correspondence is set up.

\subsubsection{Compensatory filtering in distillation set}\label{distil_temp}

We want the distillation set to be supplementary to the true structure set. Therefore the distillation set is filtered by its similarity to the PDB database, and only those which are dissimilar to the PDB sequences are kept. We selected the PDB70 sequence database as it is commonly used in template searching and the sequence similarity is sufficiently high to cover sequences in the PDB database.

"sum\_probs" is a evaluation score generated by HHsearch for each template of a query sequence, and is commonly used to rank templates. This score is related to the query sequence length, so we defined the normalized template score for each template as “sum\_probs/query\_length” for statistical purposes. For UniRef50\cite{uniref50} sequences after diversity, length, and depth filter, histograms were separately plotted for four query length intervals including [128, 256], [256, 384], [384, 512], and [512, 1024], to show the distribution of the highest template score in the .hhr files. All histograms showed apparent bimodal distributions. An empirical cutoff 0.4 was chosen to separate the distribution into two parts.  The maximum score lower than 0.4 means that even the best template is not similar enough to the query. We searched all 4M sequences and about 800k remains with highest template score < 0.4. 

\begin{figure}
\centering
\includegraphics[width=\linewidth]{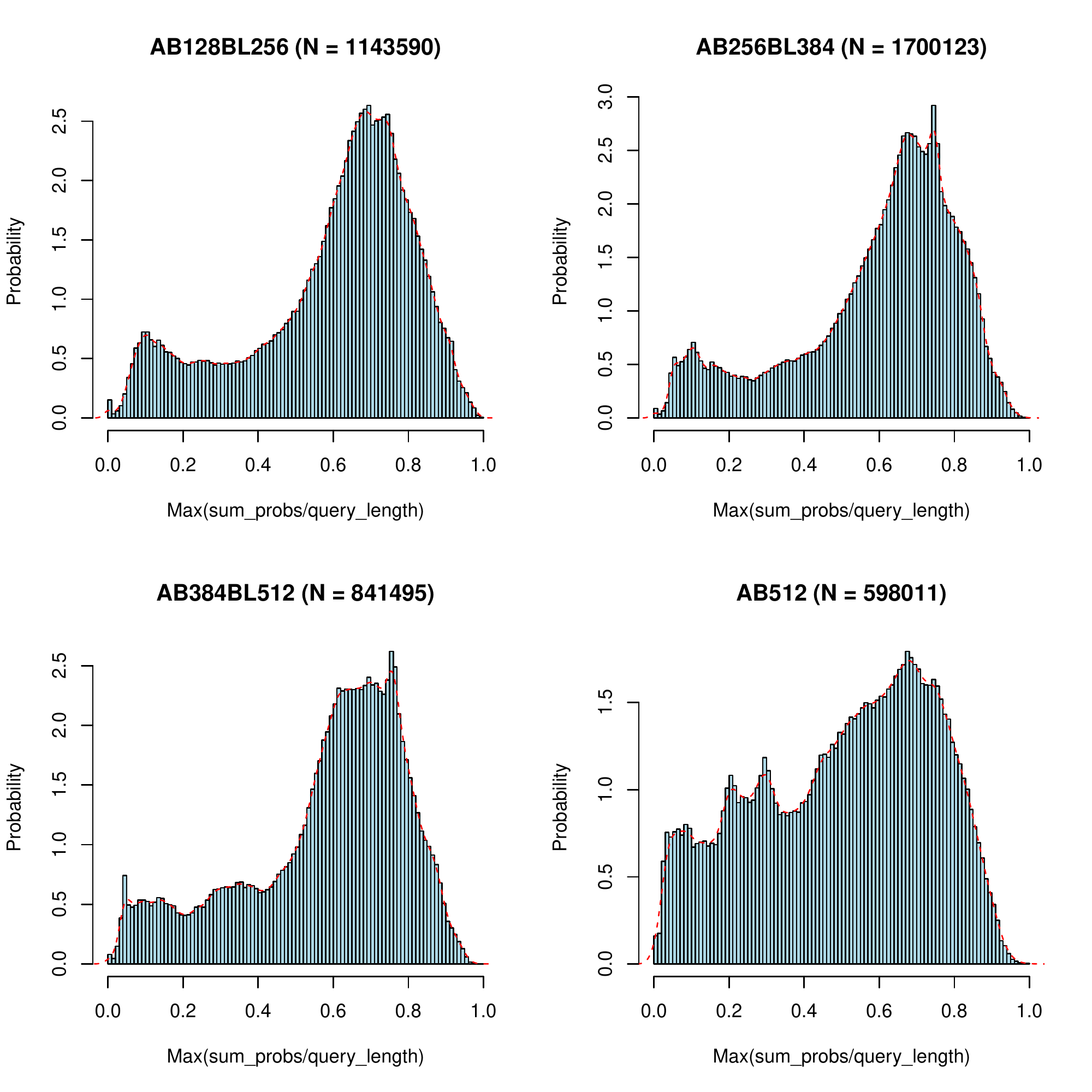}
\captionsetup{justification=centering}
\caption{Distribution of the highest template score for samples of different length group ([128, 256], [256, 384], [384, 512], and [512, 1024]).}
\label{fig:template}

\end{figure}

\subsubsection{Details of MSA, template, and structure generation}\label{gen_detail}

\subsubsubsection{MSA search with MMseqs2}\label{msa_gen}

For true structure set and distillation set, the processed sequences are treated as query sequence for MMseqs2\cite{mmseqs2} search. The core searching procedure based on the bash code from \href{https://github.com/sokrypton/ColabFold/blob/main/colabfold_search.sh}{colabfold}\cite{colabfold}. We used datasets from the uniref30\_2103.tar.gz and colabfold\_envdb\_202108.tar.gz as target database (https://colabfold.mmseqs.com/). All sequences were grouped to 2000 sequence in each mini-batch and searched in batch. Results of each query sequence from different target database were then merged and filtered by HHfiter\cite{hhsuite3} (HHfilter –id99 –v 1).

\subsubsubsection{Template search with HHsearch}\label{temp_gen}

The MSA files obtained in MMseqs search against Uniref30\cite{uniclust30} and HHblits\cite{hhsuite3} against Uniref30 were used to search PDB70 using HHSearch to get .hhr files for true structure set and distillation set, respectively. The slight difference in initial searching profile is because of differences in protocols. The template search in true structure dataset is only used to obtain similar structures for each sample, and is conducted by end of the whole protocol. While in distillation set, the template search is used to distinguish sequence that are not similar to any sequences with solved structures, and therefore is conducted before massive MSA searching. We set all parameters to default except  that “maxseq”  was set to 1000000. 

\subsubsubsection{Structure prediction with AlphaFold}\label{struct_gen}

In distillation set, we generated predicted structures for each processed sequence before confidence filter. Each structure is predicted with MSA from MMseqs2 (as described in \ref{msa_gen}). Since the filtered sequences lack confident structure templates, faked template is applied in structure prediction with code snippet from ColabFold. We used model\_1\_ptm of AlphaFold\cite{alphafold} v2.0.1 and parameters from alphafold\_params\_2021-07-14.tar.

\subsubsection{PSP lite}\label{psp_lite}

In order to increase the data accessibility and facilitate the deployment, we additionally curated a light-weight version of the full database (named as PSP lite). In this lite version, we reduced significantly (80\%) the size of the raw database while kept as much valuable information as possible. Samples in PSP lite differ with their counterparts in original PSP dataset only in MSA-relevant features. For each query sequence (or each data sample), we first filtered its MSA according to the three primary rules:

\begin{itemize}
    \item remove all MSA's with coverage lower than 50\%
    \item remove all MSA's with >90\% identity to query sequence
    \item remove all MSA's with <20\% identity to query sequence
\end{itemize}

To further reduce the size of the remaining data, we also limited the maximum MSA depth of each sample to be 512. For any sample which contains more than 512 MSA sequences after primary filtering, we selected representative MSA sequences via a heuristic strategy as follows: We initialized an MSA sequence pool using the query sequence alone, then added to this pool a new MSA sequence given that this candidate is of no more than 90\% identity to all MSAs that already in the pool, and that this candidate is closest to the query in terms of the Hamming's distance. This iterative selection stops when no more candidates can be accepted or the MSA pool is full (up to 512).

After filtering, we removed samples which contain less than 128 MSA sequences, thus guaranteeing a reasonable MSA depth for the entire database. Finally, we obtained a mini database consisting of 440k true structure samples(185GB) and 640k distillation samples in (167GB), striking a good balance between being informative and accessible.

\subsection{Statistics of Dataset}

\begin{figure}
\centering

\begin{subfigure}{0.8\textwidth}
\centering
\includegraphics[width=\linewidth]{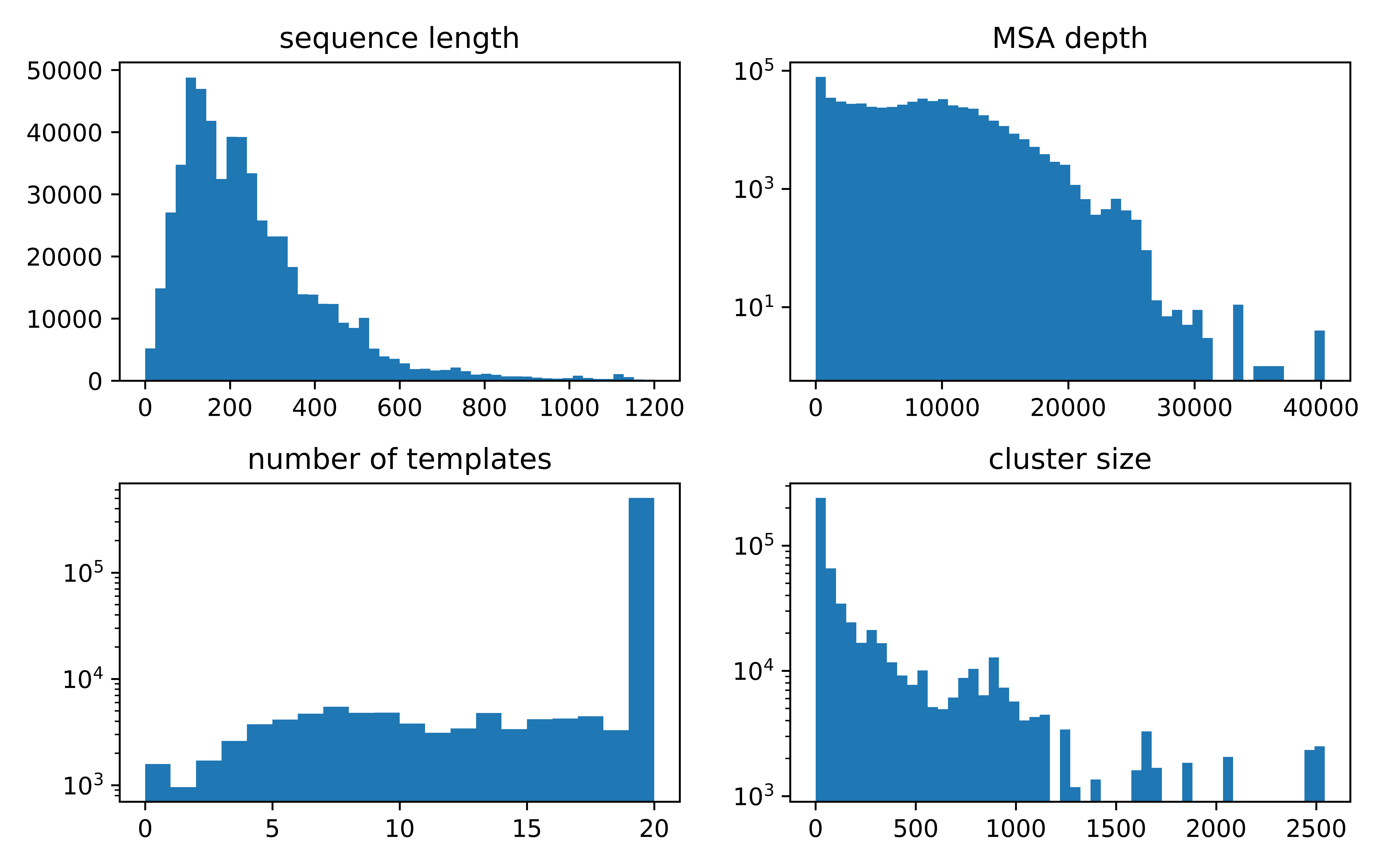}
\caption{}
\label{fig:distillation_statistics}
\end{subfigure}

\begin{subfigure}{0.8\textwidth}
\centering
\includegraphics[width=\linewidth]{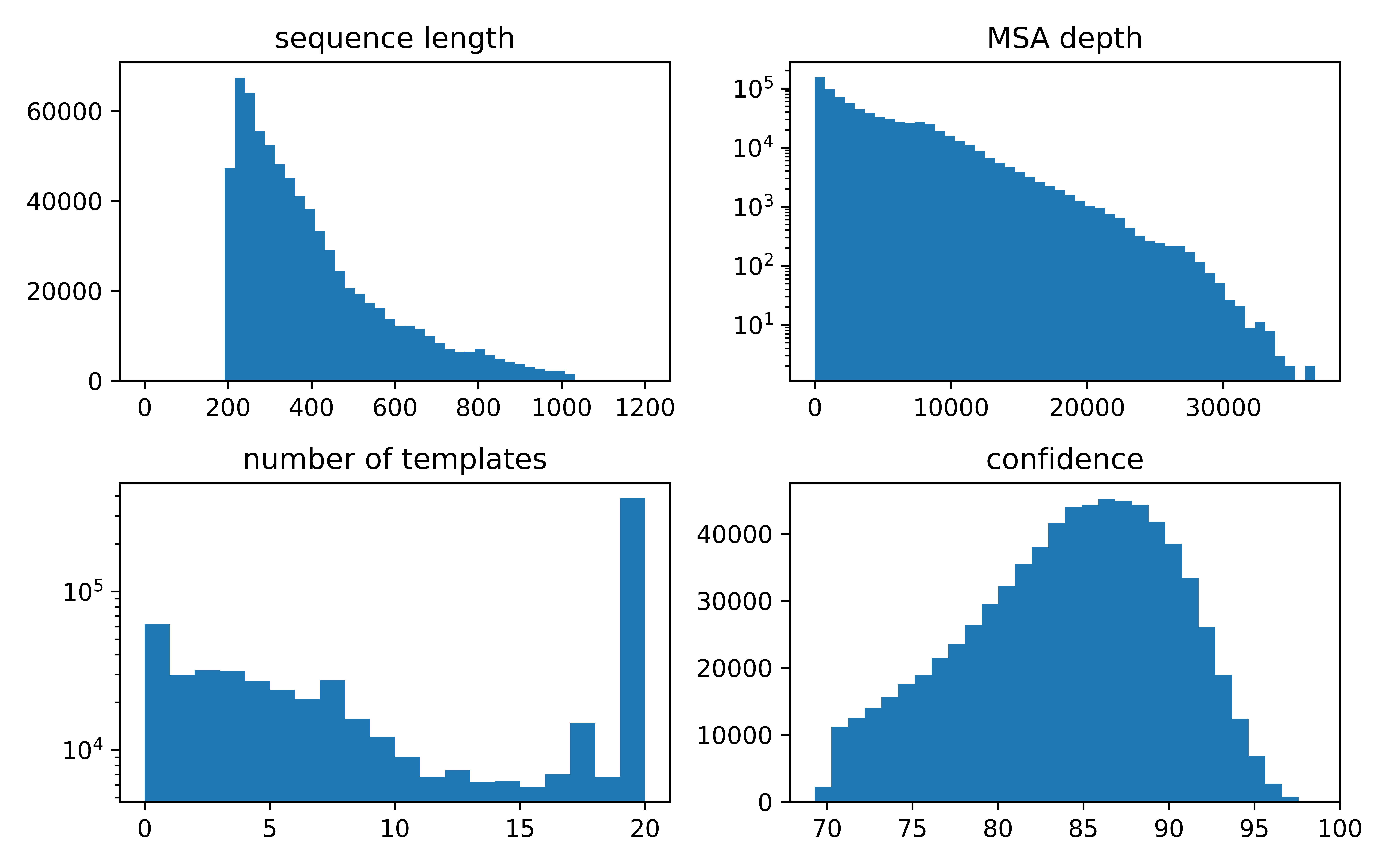}
\caption{}
\label{fig:true_structure_data_statistics}
\end{subfigure}

\caption{Statistics of true structure dataset(a) and distillation dataset(b).}
\label{fig:statistics}
\end{figure}

The statistics of the PSP dataset (true structure dataset and distillation dataset) can be found in figure \ref{fig:statistics} . Each dataset is accompanied by a json file where more detailed information for each sequence can be found: json file for the true structure dataset contains the sequence\_length, MSA depth, number of templates, cluster size, release date of the structure, resolution and the experimental method. For the distillation dataset, the json file contains sequence length, MSA depth, number of templates and the predicted confidence of structure.

\subsection{Details for Benchmark}

\subsubsection{Two-stage training evaluation}

\begin{figure}
\centering
\includegraphics[width=0.95\linewidth]{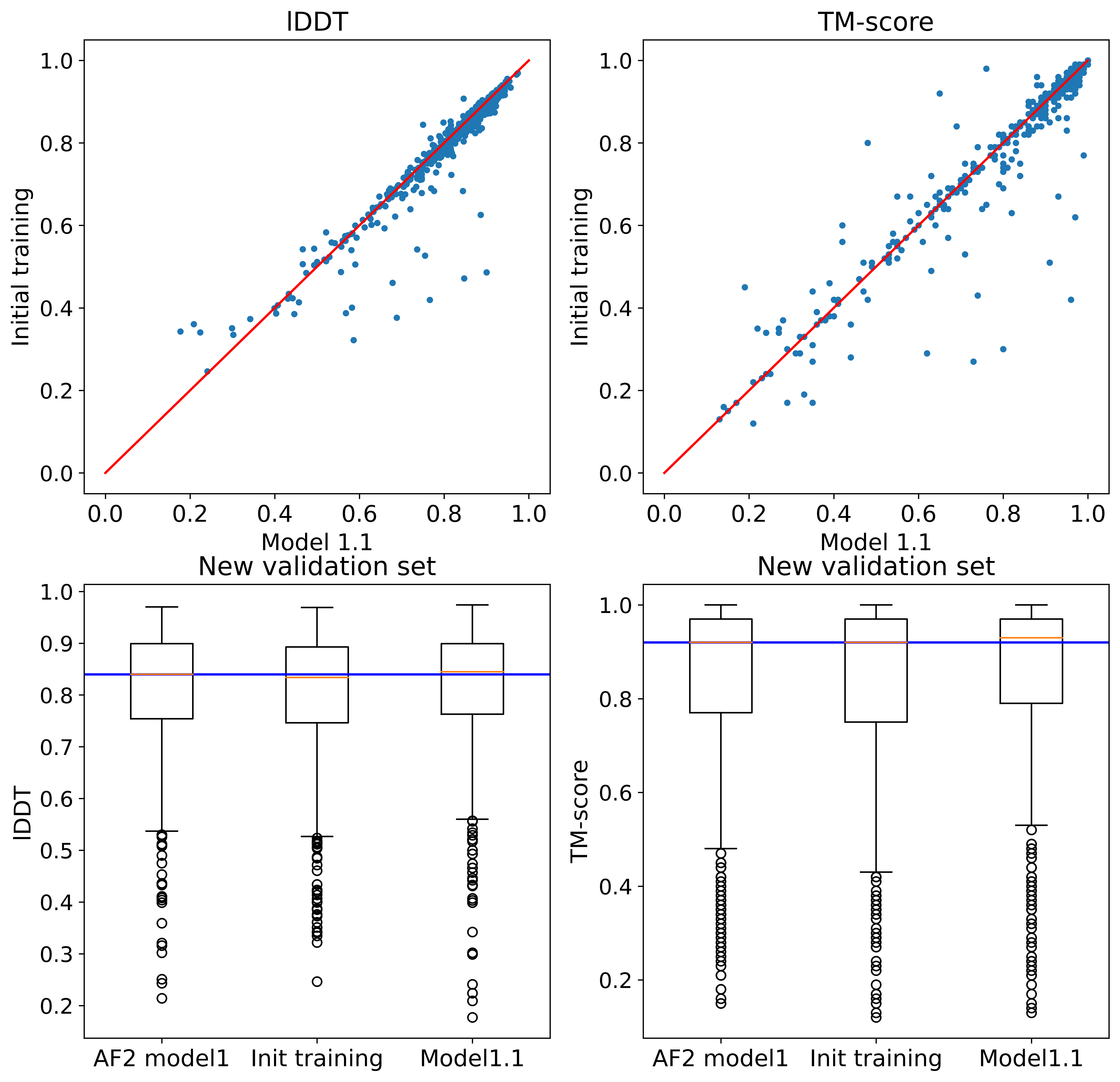}

\caption{Model evaluation. Upper figures are per-sample lDDT (left) and TM-score (right) between initial training and fine-tune. Lower figures are lDDT (left) and TM-score (right) comparison between standard AlphaFold, initial benchmark, and fine-tuned model (model 1.1).}
\end{figure}

To benchmark the effect of initial training and fine-tuning, we evaluated the model performance after initial training and fine-tune stage on the new validation dataset, respectively. It can be seen that despite the model performance is similar to that of AlphaFold (which goes through two training stages) after initial training, fine-tuning helps to fix both all-atom lDDT accuracy and C-alpha TM-scores.

\subsubsection{Details for CAMEO contest}

Summary of training protocol for CAMEO \cite{cameo1, cameo2, cameo3, cameo4} models can be found in \ref{table:training_params}. We performed the initial training with a sequence crop size 256. Then we increased the sequence crop size to 384 for the fine-tuning for which we also enabled violation losses and reduced the learning rate, resulting in two models 1.1 (with template) and 1.2 (without template). The main difference between the two models is that model 1.2 does not use the template information in fine-tune stage. 

Sergey et al. \cite{sergey_score} proposed that the potential function learnt by Alphafold2 model can score the quality of candidate protein structures without co-evolutionary information. In this method, the alternative structure is used as a template, and the sequence is used as input MSA for the model. The structures are then ranked according to the confidence and TM-score of the model output. We apply this method to score protein structure predictions predicted by model 1.1 and 1.2. We submitted three results in CAMEO Assessment weekly, including predictions from model 1.1 and model 1.2, and the scoring results of these two models.

\begin{table}[h]
\renewcommand\arraystretch{1.5}
\begin{center}
\caption{Traning protocol for CAMEO models. The fine-tuning model 1.1 and 1.2 are used in the CAMEO assessment.}
\label{table:training_params}
\begin{tabular}{lcccl}
\cline{1-4}
{\textbf{Model}}                     & \textbf{Initial training} & \multicolumn{2}{c}{\textbf{Fine-tuning}} &  \\ \cline{2-4}
                                                    & \textbf{1}                & \textbf{1.1}        & \textbf{1.2}       &  \\ \cline{1-4}
\textbf{Parameters   initialized from}              & Random                    & Model 1             & Model 1            &  \\
\textbf{Number   of templates Ntempl}               & 4                         & 4                   & 0                  &  \\
\textbf{Sequence   crop size Nres}                  & 256                       & 384                 & 384                &  \\
\textbf{Number   of sequences Nseq}                 & 128                       & 128                 & 128                &  \\
\textbf{Number   of extra sequences Nextra\_seq}    & 1024                      & 1024                & 1024               &  \\
\textbf{Initial   learning rate}                    & $10^{-3}$                 & $3*10^{-4}$         & $3*10^{-4}$        &  \\
\textbf{Learning   rate linear warm-up samples}     & 128000                     & 0                   & 0                  &  \\
\textbf{Structural   violation loss weight}         & 0                         & 1.0                 & 1.0                &  \\
\textbf{Training   samples ($10^6$)}                & 9.6                       & 3.0                 & 3.0                &  \\
\textbf{Training   time}                            & 11 days                   & 7 days              & 7 days             &  \\ \cline{1-4}
\end{tabular}
\end{center}
\end{table}

During the 1-month evaluation cycle of CAMEO from March, 25, 2022 to April, 16, 2022(table \ref{table:2}), our models ranked first (model 1.1), second (scoring model) and fourth (model 1.2) respectively, surpassing all other existing models.

\begin{table}
\renewcommand\arraystretch{1.5}
\centering
\caption{Cameo-3D 1-month performance comparison (2022-03-25 - 2022-04-16) of different severs.}
\label{table:2}
\begin{tabular}{lccl} 
\specialrule{0em}{0.1pt}{0.1pt}
\cline{1-3}
Sever Name      & Average lDDT & Average TM-score &   \\ 
\cline{1-3}
\pmb{ours(model 1.1)}        & \pmb{87.3}         & \pmb{90.6}              &   \\
\pmb{ours(quality scoring)}        & \pmb{87.1}         & \pmb{90.3}              &   \\
HeliXonAI       & 86.6         & 89.3                &   \\
SADA     & 86.5         & 88.8              &   \\
\pmb{ours(model 1.2)}        & \pmb{86.3}         & \pmb{89.4}              &   \\
pureAF2\_notemp & 86.2         & 88.6              &   \\
pureAF2\_orig   & 79.4         & 81.6              &   \\
RoseTTAFold   & 69.4         & 77.4              &   \\
\cline{1-3}
\end{tabular}
\end{table}

For CAMEO from March, 25, 2022 to April, 16, 2022, 60 target proteins in total are evaluated, while pureAF2\_orig and RoseTTAFold lack some of submissions. As the CAMEO server calculate the sum of all validate scores on number of all target proteins, scores for these 2 servers are abnormally low for this month. For fairness, we calculated the score on intersection set of validate submissions of all listed servers, 51 sequences in total, the scores are shown in \ref{table:cameo_calibrated}. On this intersection set, our models are still the best ones.

\begin{table}
\renewcommand\arraystretch{1.5}
\centering
\caption{Calibrated CAMEO-3D 1-month performance comparison (2022-03-25 - 2022-04-16) of different severs.}
\label{table:cameo_calibrated}
\begin{tabular}{lccl} 
\specialrule{0em}{0.1pt}{0.1pt}
\cline{1-3}
Sever Name      & Average lDDT & Average TM-score &   \\ 
\cline{1-3}
\pmb{ours(model 1.1)}        & \pmb{87.2}         & \pmb{90.1}              &   \\
\pmb{ours(quality scoring)}        & \pmb{86.9}         & \pmb{89.6}              &   \\
pureAF2\_orig   & 86.8         & 89.1              &   \\
\pmb{ours(model 1.2)}        & \pmb{86.7}         & \pmb{89.6}              &   \\
pureAF2\_notemp & 86.7         & 89.0              &   \\
HeliXonAI       & 86.5         & 81.4                &   \\
SADA     & 86.4         & 88.6              &   \\
RoseTTAFold   & 74.3         & 81.9              &   \\
\cline{1-3}
\end{tabular}
\end{table}

\subsection{Data Organization}

The PSP and PSP lite dataset can be separated to 2 parts respectively, the true structure data and the distillation data which are organized in the same way and can be merged easily. Due to the huge disk storage required by our dataset, we separate true structure dataset and distillation dataset to 256 parts respectively and compress their pkl and pdb to .tar.gz format. The overall size is reduced to 1.6T after compression, 800G for true structure dataset and 800G for distillation dataset. For the same reason The PSP lite dataset is divided into 32 parts for mini-true set and mini-distillation set respectively, and the size is reduced to 352G after compression, 185G for true structure dataset and 167G for distillation dataset. The new validation set is small and hence is compressed as one file for its pkl and pdb.

Taking the true structure dataset as example, for each sequence-structure data, we provide 2 files, one is in pdb format and contains the true/predicted structure information of this sequence; another is in pkl format and can be loaded by pickle library of python as normal pkl file. This pkl file contains the sequences, template and MSA information(as a python dictionary object), which are the inputs for normal structure prediction systems. This dictionary contains following attributes and significance for each is explained, as noted $N_{seq}$ stands for number of amino acids of proteins, $N_{MSA}$ for number of sequences in MSA,  $N_{template}$ for number of templates:

\begin{itemize}
    \item aatype ($N_{seq}$, 21), amino acid types in the poly-peptide chain of protein encoded as a number following the standard residue order, 20 types of amino acid + unknown, the standard residue order of 20 amino acid types is as follows: \\
    $ Alanine(A), Arginine(R), Asparagine(N), Aspartate(D), Cysteine(C),\\
    Glutamine(Q), Glutamate(E), Glycine(G), Histidine(H), Isoleucine(I),\\
    Leucine(L), Lysine(K), Methionine(M), Phenylalanine(F), Proline(P), \\
    Serine(S), Threonine(T), Tryptophan(W), Tyrosine(Y), Valine(V)$
    
    \item between\_segment\_residues ($N_{seq}$,), whether there is a domain break. Always zero for single chains, kept to be compatible with domain datasets.
    \item domain\_name (1,) , name of the domain.
    \item residue\_index ($N_{seq}$,), index of residues in the chain.
    \item sequence (1,), original sequence in string format.
    \item msa ($N_{MSA}$, $N_{seq}$), MSA of the sequence, with number encoding 20+1 amino acids
    \item deletion\_matrix\_int ($N_{MSA}$, $N_{seq}$), numbers of deletion for MSA of the sequence
    \item num\_alignments ($N_{seq}$,), number of alignment for each residue.
    \item template\_aatype ($N_{template}$, $N_{seq}$, 22), amino acid types of template of the target sequence, we select up to 20 best template for each target sequence, the final dimension is 20 types of amino acid + unknown + gap, the amino acid follow the same order as aatype.
    \item template\_all\_atom\_positions ($N_{template}$, $N_{seq}$, 37, 3), atom positions of templates, here we follow the format of atom37 in AlphaFold2 \cite{alphafold}.
    \item template\_all\_atom\_masks ($N_{template}$, $N_{seq}$, 37), mask for atom positions of templates.
    \item template\_domain\_names ($N_{template}$,), domain name of templates.
    \item template\_e\_value ($N_{template}$, 1), e\_value of templates.
    \item template\_neff ($N_{template}$, 1), neff of templates.
    \item template\_prob\_true ($N_{template}$, 1), prob\_true of templates.
    \item template\_release\_date ($N_{template}$,), release date of templates.
    \item template\_score ($N_{template}$, 1), score of templates.
    \item template\_similarity ($N_{template}$, 1), similarity of templates.
    \item template\_sequence ($N_{template}$,), sequence of templates in string format.
    \item template\_sum\_probs ($N_{template}$, 1), sum\_probs of templates.
    \item template\_confidence\_scores ($N_{template}$, $N_{seq}$), confidence per residue for templates.

\end{itemize}

\newpage

\begin{sidewaystable}[!ht]
    \centering
    \caption{Details of database in use}
    \label{table:database}
    \begin{tabular}{|l|p{3cm}<{\centering}|p{8cm}<{\centering}|p{3.5cm}<{\centering}|}
    \hline
        Database & Usage & Description & Version \& Url  \\ \hline
        UniRef50\cite{uniref50} & starting database for following length, coverage, and MSA depth based filtering in distillation set & Clustered sets of sequences from UniProtKB and selected UniParc records(largest database of known sequences). This hides redundant sequences and obtains complete coverage of the sequence space.  & \href{https://www.ebi.ac.uk/training/online/courses/uniprot-quick-tour/the-uniprot-databases/uniref/}{UniRef50\_2021\_02}  \\ \hline
        UniClust30\cite{uniclust30, mmseqs2} & MSA searching database for HHblits and MMseqs in true structure dataset and distillation set & Databases generated by cluster UniProtKB sequences at the level of 30\% pairwise sequence identity. The clusterings show a high consistency of functional annotation owing to an optimised clustering pipeline that runs with our MMseqs2 software for fast and sensitive protein sequence searching and clustering. & mmseqs:\ \  \ \
        \ \href{https://colabfold.mmseqs.com}{uniref30\_2103} \ \  \ \
        HHblits: \href{https://uniclust.mmseqs.com/}{Uniclust30\_2020\_06}\   \\ \hline
        Colabfold\_envdb\cite{colabfold} & MSA searching database for MMseqs in true structure dataset and distillation set & Colabfold\_envdb is part of ColabFold databases which are MMseqs2 expandable profile databases to generate diverse multiple sequence alignments to predict protein structures.   & \href{https://colabfold.mmseqs.com/}{colabfold\_envdb\_202108}  \\ \hline
        PDB\cite{pdbdatabase} & starting database for following data cleaning in true structure set & The Protein Data Bank (PDB) is a database for the three-dimensional structural data of large biological molecules, such as proteins and nucleic acids. The data, typically obtained by X-ray crystallography, NMR spectroscopy, or, increasingly, cryo-electron microscopy & \href{https://ftp.rcsb.org/pub/pdb/data/structures/divided/mmCIF/}{PDB\_20211013}  \\ \hline
        PDB70\cite{hhsuite3} & template searching database for true structure dataset and distillation dataset & Full chains of known 3D structures while selected by PSI-BLAST alignments produced with sequences of PDB full chain representativies (<70\% sequence identity) as queries & \href{https://wwwuser.gwdg.de/~compbiol/data/hhsuite/databases/hhsuite_dbs/pdb70_from_mmcif_latest.tar.gz}{PDB70\_20211106}  \\ \hline
    \end{tabular}
\end{sidewaystable}

\end{document}